\documentclass[journal]{IEEEtran}
\usepackage{graphicx}
\begin{document}

\title{Why white noise is not enough? \\
On using radio front-end models while designing 6G PHY}

\author{Pawel Kryszkiewicz, Pawel Sroka, Marcin Hoffmann, Marcin Wachowiak}

\markboth{Pawel Kryszkiewicz, Pawel Sroka, Marcin Hoffmann, Marcin Wachowiak}
{Why white noise is not enough? On using radio front-end models while designing 6G PHY}

\maketitle

\begin{abstract}
From generation to generation there are increasing requirements for wireless standards both in terms of spectral and energy efficiency. While up to now the layered wireless transceiver architecture worked allowing for, e.g., separation of channel decoding algorithms from front-end design, this may need reconsideration in the 6G era. Especially the hardware-originated distortions have to be taken into account while designing other layer algorithms as the high throughput and energy efficiency requirements will push these devices to their limit revealing their nonlinear characteristics. This position paper will shed some light on new degrees of freedom while cross-layer designing and controlling multicarrier and multiantenna transceivers of 6G systems.
\end {abstract}


\section{Introduction}
\begin{figure}[htb]
    \centering
    \includegraphics[width=3.5in]{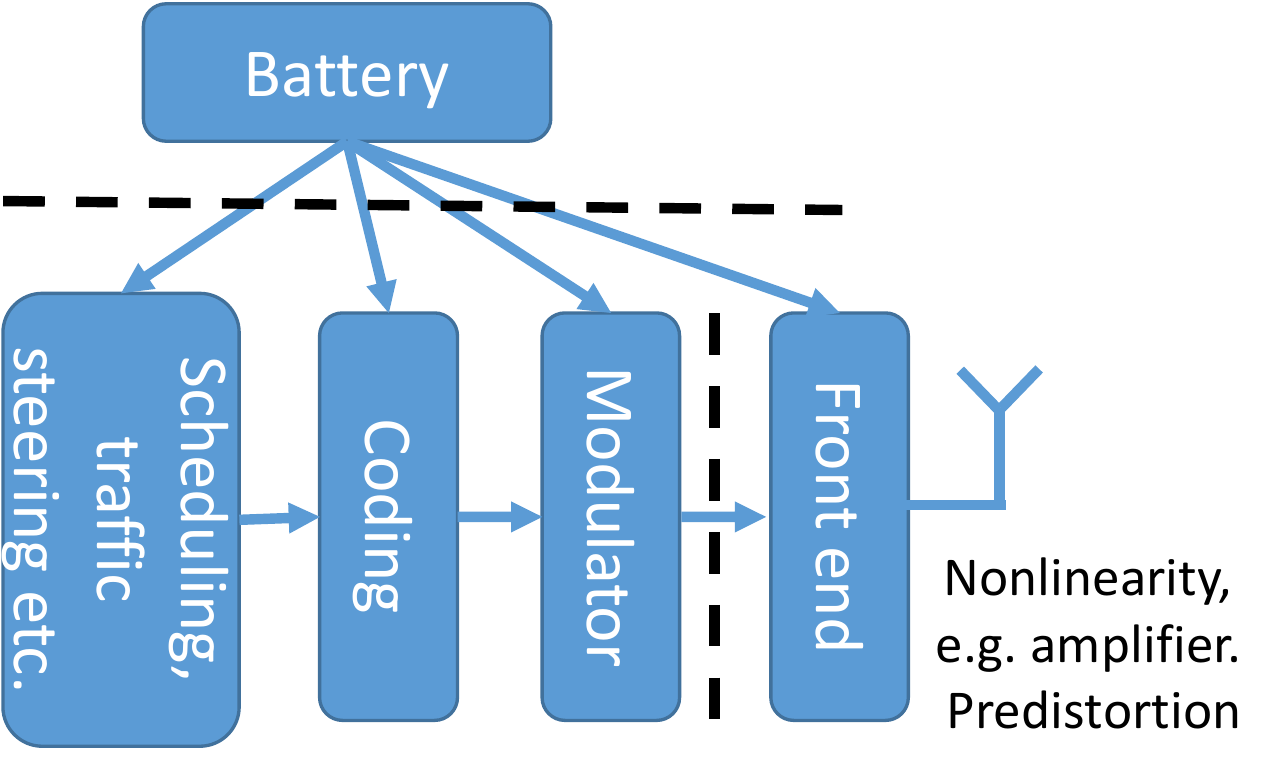}
    \caption{Diagram of a typical wireless transmitter. Dashed lines show typical splits between design domains.}
    \label{fig:diagram}
\end{figure}

Increasing number of users and applications results in more demanding requirements on wireless networks. This can be met with the 5G network employing simultaneous densification of base stations, spectral efficiency improvement techniques, and an increase in the utilized bandwidth. However, this comes at the cost of increased energy consumption of wireless transceivers resulting, e.g., from a higher number of radio front-ends (antennas) and more complex signal processing both at the transmitter and the receiver side. Therefore, the next, 6G technology will need new design paradigms to reflect the rising needs and new constraints. Up to now, the effort was on obtaining maximal spectral efficiency for a "perfect" radio front-end. However, while pushing the transceiver design into its spectral and energy consumption limits, new phenomena become significant and need to be taken into account. This can be analyzed using a typical wireless transmitter diagram in Fig. \ref{fig:diagram}. In most cases each of the depicted blocks, e.g., modulator or battery, is designed independently. The dashed line shows typical splits between design domains. Most researchers, e.g., in the field of radio resources management, assume perfect and linear modeling of electronic components constituting the radio front-end, e.g., Power Amplifiers (PAs). In reality, a linear increase of mean transmit power results in a linear increase of consumed power and no signal distortion only when a high back-off (difference between clipping level and mean transmit power) is used. Unfortunately, this front-end model cannot be used for designing highly energy-efficient systems. For this case of a wireless system operating close to its energy efficiency maximum, a nonlinear front-end characteristic and its influence on many system metrics have to be considered, e.g., for a given PA the increased transmission power results in increased self-interference power generated both in the transmission band and outside the transmission band, widening the system bandwidth and possibly decreasing spectral efficiency\cite{gharaibeh2011nonlinear}. At the same time, the transmitted signal, as a result of PA nonlinear characteristics, has its distribution changed, resulting in a nonlinear increase in power consumption\cite{Ochiai_TCOM_2013_PA}. These problems with nonlinear behavior can be extended to some other components of the wireless transceiver, e.g., Analog-Digital Converters (ADCs) or batteries that power the front-end. As such, nonlinear front-end models should be considered while designing specific solutions for wireless transceivers and systems. This paper presents a couple of new approaches that can be utilized in multiple components of a future 6G system. 

\section{Optimization of PA operating point for the OFDM signal}
First, the operating point of a PA transmitting Orthogonal Frequency Division Multiplexing (OFDM) waveform can be optimized in order to maximize the total link spectral or energy efficiency\cite{Kryszkiewicz_Sensors_2023_Battery}. Commonly, the operating point is adjusted in order to achieve a given spectral emission mask (for out-of-band frequency region) or Error Vector Magnitude (EVM) (in the in-band region) at the transmitter output. However, if the operating point of the PA is adjusted in order to maximize the link's spectral efficiency, significant gains in comparison to contemporary solutions are possible. Most importantly, this requires precise modeling of the transmitted waveform (in most cases its amplitude distribution is enough) and the front-end nonlinearity in addition to the standard phenomena like multipath fading or interference observed at the receiver. If the goal function changes to maximization of energy efficiency, in addition, models of energy consumption are needed. This obviously depends on a given implementation, e.g., PA class, or utilized modules, e.g., what kind of coding is used. While energy efficiency is the most important aspect for battery-powered devices, battery models are typically omitted while designing communications systems. In reality, a battery can reveal significant nonlinear characteristics, e.g., the so-called "rate-capacity" effect causes two times higher instantaneous power required to power the power amplifier, which results in a more than two-fold increase in the energy drained from the battery. These all models are to be combined together to find the optimal operating point of the PA.

\begin{figure}[htb]
    \centering
    \includegraphics[width=3.4in]{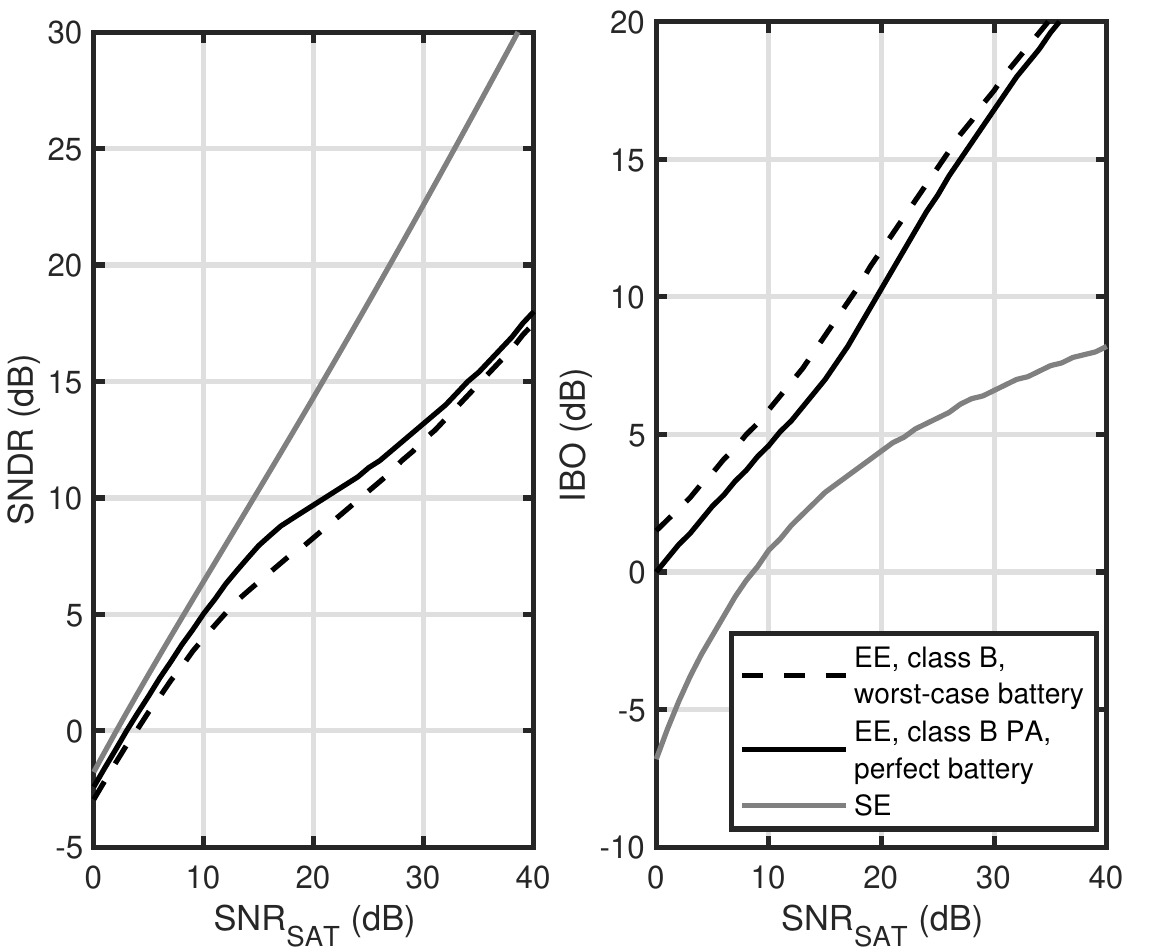}
    \caption{Optimal values of Signal to Noise and Distortion Ratio (SNDR) and Input Back-Off (IBO) as a function of saturation Signal-to-Noise Ratio (SNR) while Spectral Efficiency (SE) or Energy Efficiency (EE) maximization for two battery models.}
    \label{fig:SNDR_IBO_SNR_SAT}
\end{figure}

An example solution, based on mathematical framework provided in \cite{Kryszkiewicz_Sensors_2023_Battery}, is shown in Fig. \ref{fig:SNDR_IBO_SNR_SAT}. It is assumed that the transmitter uses soft-limiter, class B PA. The operating point of the PA, i.e., IBO, is optimized in order to achieve maximal spectral efficiency (SE) or energy efficiency (EE). Moreover, two different battery models are considered. The maximal unclipped PA output is assumed to be 100 mW and the constant power consumption of other electronic components in the transmitter is set to 10 mW. The $SNR_{\mathrm{SAT}}$ is maximal signal-to-noise ration achievable over given link, i.e., when transmitting a single carrier of power 100 mW. As expected the SE maximization achieves the highest Signal to Noise and Distortion Ratio (SNDR) in the whole  SNR range. Most importantly, the optimal IBO value changes significantly depending on the channel conditions. If the EE is to be optimized there is no point in achieving so high SNDR (as it results in higher energy consumption) resulting both in different SNDR and optimal IBO curves. Finally, it is visible that a battery model should be considered while optimizing EE of a transmitter as it has influence both on SNDR and optimal IBO values.

\section{Front-end characteristics enabling optimization of non-Gaussian waveform}
While in most OFDM applications a large number of occupied subcarriers is considered central limit theorem can be used to model instantaneous OFDM signal amplitude using Rayleigh distribution, e.g., in the use case mentioned above. However, if a small number of subcarriers is used, e.g., in the uplink from some sensor devices, the above-mentioned distribution is not valid anymore. Such a system, relying on orthogonal frequency-division multiple access (OFDMA) is considered in future 5G and 6G systems in the scope of massive machine-type communications (mMTC), where large numbers of simple devices can connect to a base station to transmit small amounts of data. The optimization of the PA operating point together with the number of utilized subcarriers, influencing complex envelope power variation, constitutes additional degrees of freedom in such a sensor network optimization. The more subcarriers are used the higher instantaneous power variations are expected, resulting in a higher amount of nonlinear distortion. At the same time, a wider spectrum allows typically for higher throughput. As such optimization of the operating point for such a scenario is not a trivial task. Moreover, the frequency-domain characteristic of nonlinear distortion can be used while designing subcarriers allocation procedures. While the OFDM nature guarantees that subcarriers are orthogonal, the spectral regrowth caused by nonlinear PA results in users utilizing adjacent frequency channels causing distortion to one another. This noise-like effect will appear contributing to the multi-user interference, with the interference level depending on the distance in frequency from the subcarriers used by the interferer. From the resource allocation point of view, such additional interference will play a significant role in the allocation of subcarriers to different users, as it may significantly degrade the quality of transmission in adjacent sets of resources. Thus, alternatives in the form of reducing the number of allocated subcarriers, which typically results in lower distortion level or improved link budget, or introducing some null subcarriers in the form of guard bands, should be considered.
    
\begin{figure}[htb]
    \centering
    \includegraphics[width=3.4in]{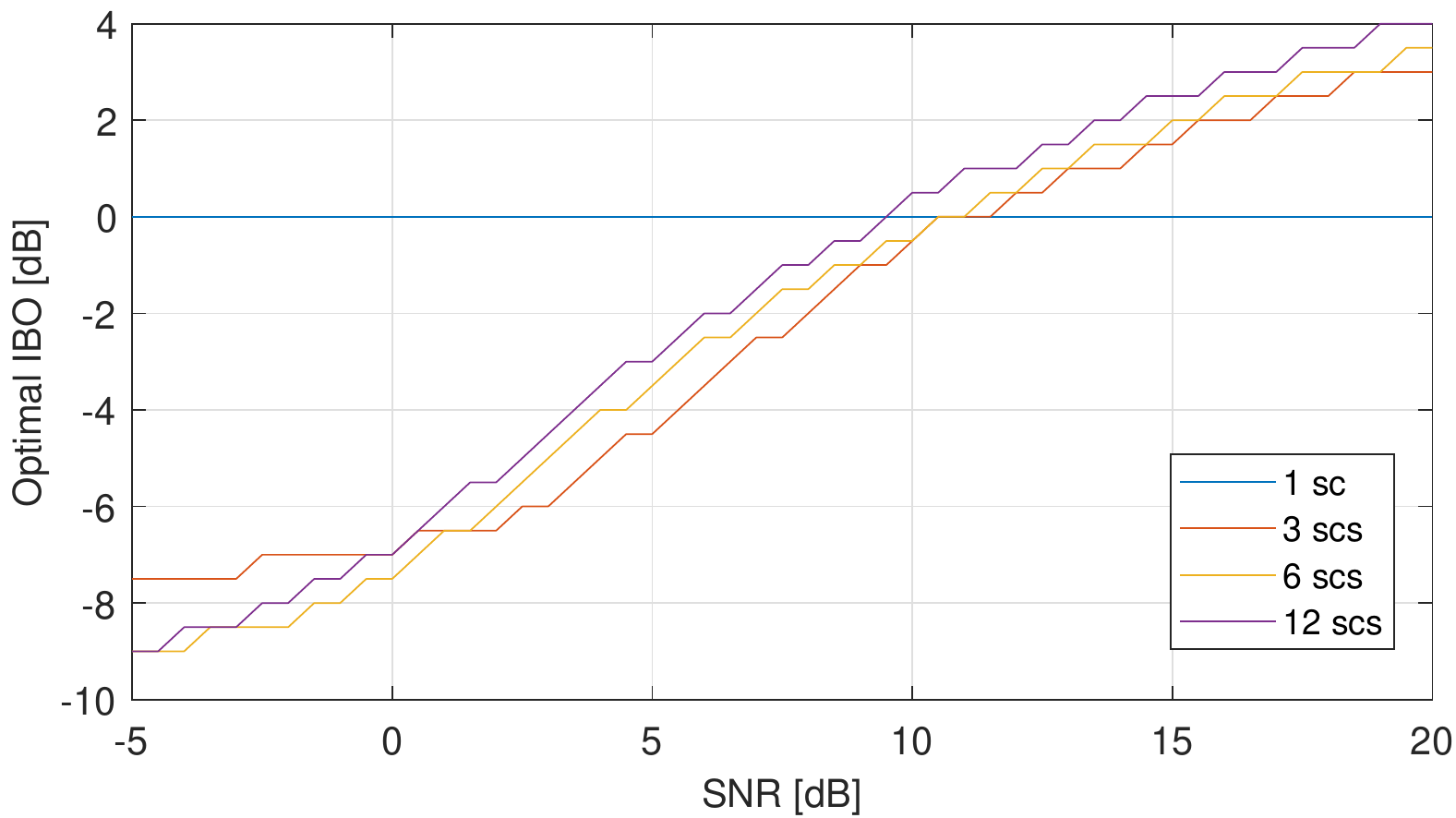}
    \caption{Optimal operation point of the PA (in terms of data rate maximization) vs. the expected SNR with low number of subcarriers used in OFDM transmission.}
    \label{fig:ibo_vs_scs}
\end{figure}

    As an example we show in Fig. \ref{fig:ibo_vs_scs} the optimal, i.e., maximizing data rate, Input Back Off (IBO) of the PA as a function of link's Signal to Noise Ratio (SNR) and number of utilized subcarriers (from 1 to 12). The clipper/soft-limiter PA model is used. IBO is a ratio between the maximal unclipped input power of the amplifier and the mean input signal power. For the single carrier transmission it is reasonable to transmit signal with maximal possible power without introducing any distortion, i.e., 0 dB IBO as the signal has constant envelope. For higher number of subcarriers it is visible that in general optimal IBO is the higher the higher is the SNR. Usage of low IBO in this case will result in the higher wanted signal power but significantly increased distortion power at the receiver. Most importantly, the curves differ between different number of subcarriers showing there is a new degree of freedom in optimization of IoT devices.

\section{Front-end aware Massive MIMO system} 
\begin{figure}[htb]
    \centering
    \includegraphics[width=3.5in, trim={0 0.5cm 0 1.4cm},clip]{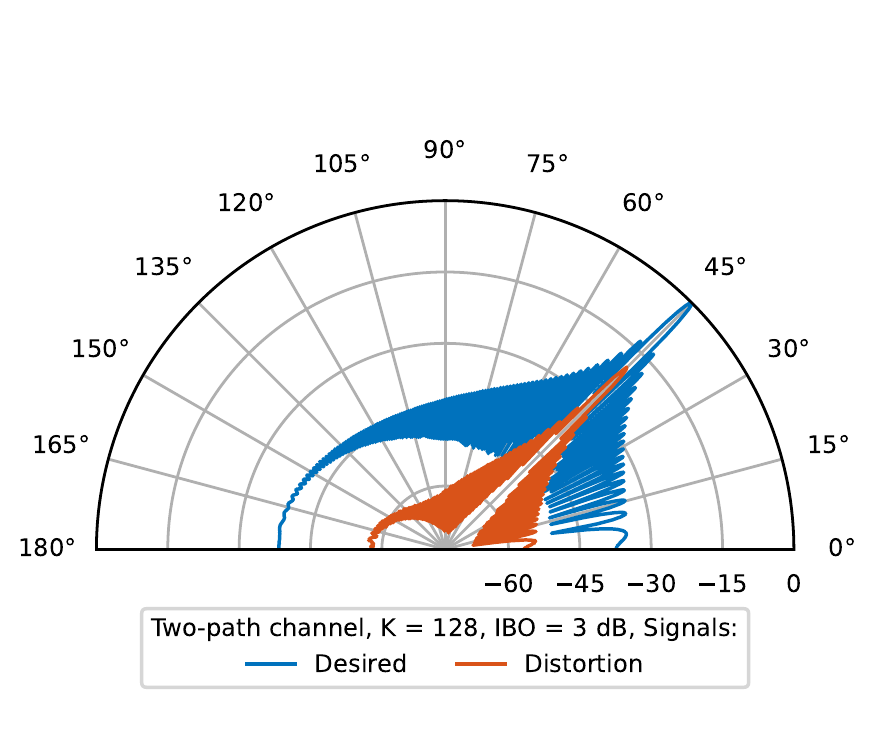}
    \caption{Normalized radiation pattern of desired and distortion signal components in regard to azimuth angle for two-path channel model, MRT precoding for azimuth $45^\circ$, 128 antennas and IBO = 3 dB.}
    \label{fig:2path_desired_vs_distortion_beampattern}
\end{figure}

Massive MIMO (mMIMO) is currently at the heart of the fifth generation of wireless networks and is envisioned to be employed also in future generations. While it is commonly combined with OFDM, it inherits a similar nonlinearity response. However, now, in addition to spectral regrowth, the unequal spatial distribution of the nonlinear distortion is revealed. The nature of mMIMO causes the signals transmitted from each array's antenna to add, e.g., in-phase boosting the received power in a given location or in a counter-phase causing signals fading, depending on the utilized precoding. Now, when the nonlinearities, generated and transmitted by each component front-end, come into play, their addition has to be considered as well. While there was a belief that nonlinear distortions will distribute omnidirectionally \cite{Bjorson_2014_EE}, there are some cases, e.g., with line-of-sight propagation or a low number of layers, when the nonlinear distortions have the same spatial distribution as the wanted signal \cite{Mollén_OOB_MIMO}.
An example is shown in Fig. \ref{fig:2path_desired_vs_distortion_beampattern}. This depicts normalized radiation pattern of an OFDM waveform radiated from 128 transmitters. In each transmitter the same PA model is used, i.e., soft-limiter of 3 dB IBO. The Maximal Ratio Transmission (MRT) precoding for the azimuth of $45^\circ$ is assumed. It is visible that both the desired signal and distortion have similar spatial characteristics.
In other words, in this case, the increase in the number of antennas does not influence the Signal-to-Distortion Ratio (SDR). Obviously, this depends on the wireless channel properties. The SDR value at the scheduled receiver for 3 different number of antennas, 2 types of channels\cite{3gpp_nr_channel_model} and varying IBO is depicted in Fig. \ref{fig:SNDR_IBO_SNR_SAT}. As expected in all the cases the SDR rises with IBO. Most important for the Massive MIMO case is the array gain, i.e., if full array gain is available only for wanted signal it is expected that SDR rises by around 12 dB by changing 1 antenna to 16. However, while it reaches around 9 dB for NLOS channel, it is smaller than 5 dB for LOS case. This shows that increasing the number of antennas will not solve the Massive MIMO nonlinear front-end problem. 
\begin{figure}[htb]
    \centering
    \includegraphics[width=3.5in]{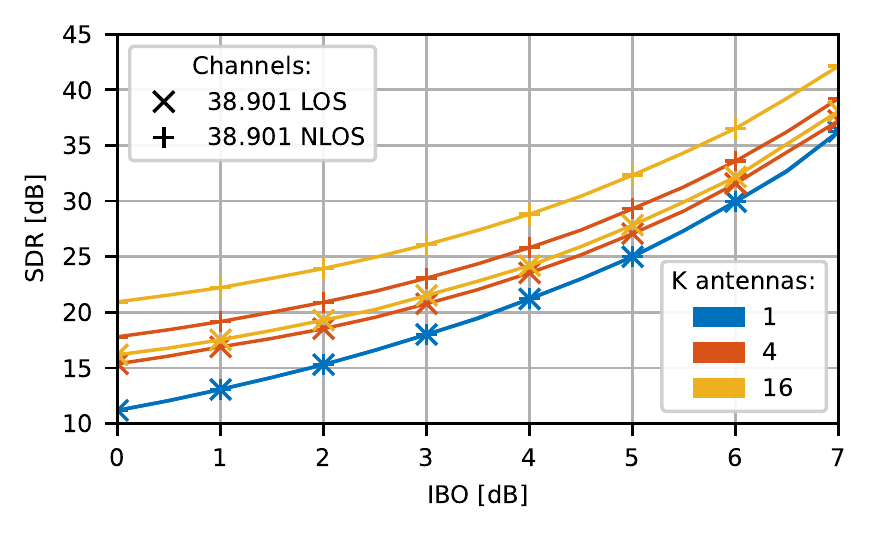}
    \caption{SDR as a function of IBO for 3GPP channel models, and varying number of antennas. MRT precoding for $45^\circ$ azimuth angle applied.}
    \label{fig:ber_out_vs_berin_csi}
\end{figure}

Even though massive MIMO can combat, e.g., high path loss or multi-user interference, nonlinear distortion remains a valid problem that has to be addressed. Obviously, this can be done using some OFDM-typical methods, such as Peak to Average Power Ratio (PAPR) reduction algorithms \cite{Kryszkiewicz_ACTR_2018} or front-end predistortion. However, it is also possible to utilize some mMIMO-specific solutions, e.g., precoding that reduces nonlinearity\cite{zayani2019_MIMO_precoding}. While the method in \cite{zayani2019_MIMO_precoding} is designed for a single carrier system, mMIMO precoding aware of OFDM processing is still missing. Obviously, the nonlinearity problem can be addressed also at the other link side, i.e., by utilizing advanced receivers. It has been shown in \cite{Zhidkov_2019_belief_propagation} for a single antenna OFDM system, that nonlinear distortion can be treated as a kind of redundancy coding, distributing information from a given subcarrier to other subcarriers, thus increasing frequency diversity. The extension of this work to an mMIMO system is an open topic as well.

\section{Front-end modeling for ML-based network optimization}
The utilization of front-end characteristics can go far beyond the link-level optimization, providing new degrees of freedom. The existence of nonlinearity should be considered while analyzing the 5G and beyond wireless systems on a system level. In this context, especially if this happens in a complex system composed of multiple transmit antennas, multiple base stations that interfere, and utilizing multi-stage processing, e.g., some proprietary traffic steering or scheduling algorithms, control and optimization of such a system is a complex task. In such a case analytical models do not exist or are significantly simplified, and we cannot rely on standard optimization methods. Though the 6G systems are expected to move toward the utilization of Machine Learning (ML) and Artificial Intelligence (AI) in order to optimize network parameters based on the analysis of real-data \cite{Yang_2020_6Gintelligence}. In this context, the most straightforward approach is to optimize the operating point of the PAs used at base stations. This parameter can be used to find a balance between the transmitted power, power amplifier efficiency, and nonlinear distortion level. Obviously, various goal functions can be used, e.g., to maximize spectral efficiency, improve the cell-edge users throughput, or maximize energy efficiency. As a reference, a classic method is proposed in~\cite{Tavares_2016_IBO_ref}, where the  optimization of PA operation point is considered, aiming at achieving maximal signal-to-noise and distortion power ratio in a single OFDM link. While the authors proposed an analytical solution, the considered system model was relatively simple, e.g., only flat fading was considered. This is one of the reasons why this solution can be not considered optimal in many situations. Therefore, we propose to utilize some Reinforced Learning technique, where, depending on the network state (which can be modeled in various ways, initially we resort to users' path-losses), various PA operating points are considered, and via the network's throughput observation the optimal operating point is selected.

\begin{figure}[htb]
    \centering
    \includegraphics[width=3.3in]{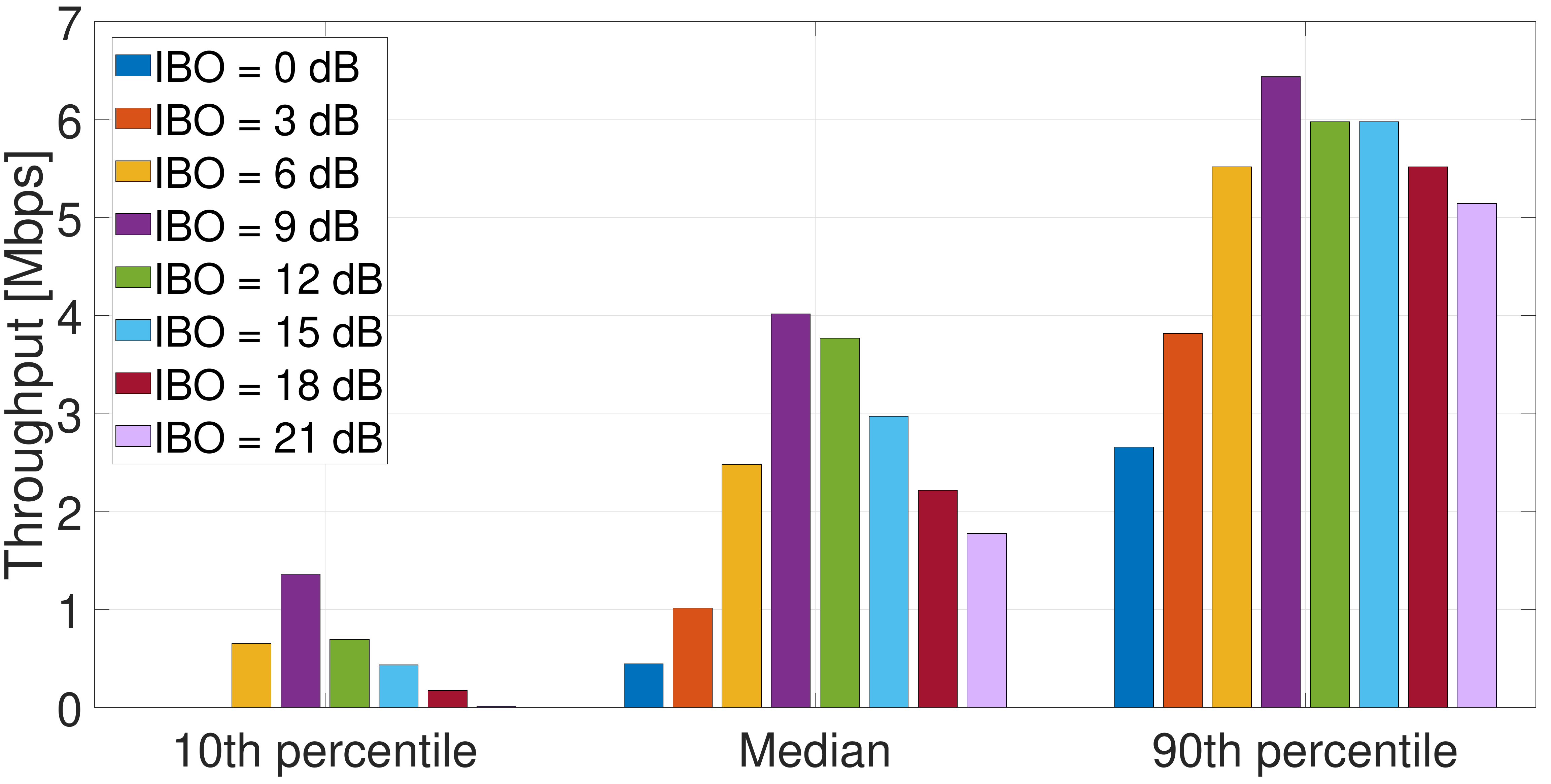}
    \caption{Throughput of cell-edge (10th percentile), median, and the best (90th percentiles) users obtained over the  training epoch for different values of IBO.}
    \label{fig:ibo_rate_bar}
\end{figure}

We have evaluated the proposed solution in terms of computer simulations. In the simulation scenario, we have considered downlink in a single mMIMO cell with one BS equipped with a rectangular antenna array of 128 elements and operating at a center frequency of $3.6$~GHz with a bandwidth of $25$~MHz. Due to the low complexity of both hardware and signal processing, we have considered analog beamforming with a so-called Equal Gain Transmission (EGT) precoder~\cite{zhang2014}. This type of precoder ensures that equal power is being allocated per BS's antenna. The PA installed at the BS along with its non-linear effects is modeled as a soft limiter \cite{gharaibeh2011nonlinear}. We used an Wireless Insite\textregistered \hspace{0.1cm} Ray-Tracing software to generate accurate radio channel coefficients for 150 sets of 20 UE distributed uniformly over the cell area. The single set is named step, while all sets constitute epoch. Throughout the multiple epochs, the RL agent was applying different values of IBO (constant over the whole epoch), and related user throughputs were observed as rewards. The related results are depicted in Fig.~\ref{fig:ibo_rate_bar}
In the figure, it is clearly visible that there exists optimal value of IBO$=9$~dB for cell-edge users (10th percentile), median users, and the best users (90th percentile). This value provides the best balance between the users' received power and non-linear distortion. Choosing IBO, which is too close to the PA saturation point ($0$~dB, $3$~dB, $6$~dB) reduces the user throughput due to the high non-linear distortion level. On the other hand, for high values of IBO, the received power is too small. This phenomenon is especially visible for the cell edge (10th percentile) of users. From this perspective ML models can be further trained to adjust IBO in a context-dependent manner, i.e., based on the users' path-loss distribution, or possible location information.

\section{Conclusions}
The above discussion sheds some light on a typically overlooked problem of nonlinearities present in wireless transceivers. The typical mathematical models resorting to white noise and multipath fading channels cannot be used if our aim is to design an ultimately efficient 6G system. However, this requires a significant effort in realistic modeling and optimization of the wireless front-end while designing higher-layer algorithms. This will open new degrees of freedom, at the cost of increased complexity. One possibility is to support a classic convex optimization solution with machine learning tools.


\section*{Acknowledgements}
This research was funded by the Polish National Science Centre, project no. 2021/41/B/ST7/00136. For the purpose of Open Access, the author has applied a CC-BY public copyright license to any Author Accepted Manuscript (AAM) version arising from this submission. https://doi.org/10.26636/jtit.2023.170523

\bibliographystyle{IEEEtran}
\bibliography{IEEEexample}


\end{document}